\documentstyle[12pt]{article}
\newcommand{\eps}{\epsilon}
\newcommand{\del}{\delta}
\newcommand{\om}{\omega}
\newcommand{\lam}{\lambda}
\begin{document}

\title{Averaged equations for Josephson junction series arrays with
LRC load}

\author{Kurt Wiesenfeld\thanks{Supported by the Office of Naval
Research under contract N00014-91-J-1257}\\
School of Physics\\ Georgia Tech\\
Atlanta, GA 30332\\
kw2@prism.gatech.edu\\
\and
James W. Swift\thanks{Supported by an Organized Research Grant
from NAU} \\
Department of Mathematics\\ Northern Arizona University \\
Flagstaff, AZ 86011-5717\\James.Swift@nau.edu}
\date{Submitted to Phys. Rev. E on August 22, 1994}
\maketitle

\begin{abstract}
We derive the averaged equations describing a series array of
Josephson junctions shunted by a parallel inductor-resistor-capacitor
load.  We assume that the junctions have negligable capacitance
($\beta = 0$), and derive averaged equations which turn out to be
completely tractable:  in particular the stability of both in-phase and
splay states depends on a single parameter, $\del$.  We find an
explicit expression for $\delta$ in terms of the load parameters and
the bias current.  We recover (and refine) a common claim found in
the technical literature, that the in-phase state is stable for inductive
loads and unstable for capacitive loads.
\end{abstract}
\newpage

\section{Introduction}
\label{sec:intro}
Josephson junction arrays are perhaps the most widely studied class
of coupled nonlinear oscillator systems.  This stems in large part
from their relevance in a number of applications, including their use
as voltage standards\cite{ref1} and their potential as sub-millimeter
wave generators\cite{Jain&al84} and parametric
amplifiers\cite{ref3}.  They
also serve as prime examples of nonlinear dynamical systems with
many degrees of freedom.  Particularly good progress has been made
for a subclass of this category, namely globally coupled oscillators.
Examples of this type --- where each oscillator is coupled with equal
strength to all others --- arise not only in the context of electrical
circuits, but in the fields of laser physics and classical mechanics as
well.

Recent theoretical work has shown that some Josephson arrays have
remarkable dynamical properties.  The most striking discovery was
made by Watanabe and Strogatz~\cite{Wat&Str94} for the class of
arrays depicted in fig.~\ref{schematic}, namely one-dimensional
series arrays of $N$ identical zero-capacitance junctions, driven by a
constant current and shunted by a parallel load.  Using a clever
change of coordinates they showed that the differential equations
admit $N-3$ independent constants of motion, for any $N>3$.
Furthermore, they found a rigorous reduction of the problem to a
5-dimensional system of differential equations, independent of $N$.

The same technique allowed Watanabe and Strogatz to completely
analyze the dynamics of the $N$-oscillator system

\begin{equation}
\dot{\varphi}_i = 1 + \frac{\kappa}{N} \sum_{j=1}^N \cos(\varphi_j -
\varphi_i - \delta) ~, ~~\: i = 1, . . . , N .
\label{target}
\end{equation}
They observed in particular that the central issue --- whether the
attracting dynamics is the in-phase (i.e. synchronized) oscillation or
an incoherent state --- depends only on the sign of $\kappa \sin
\delta$.

The purpose of the present paper is to derive Eq.~(\ref{target})
as the averaged version of the Josephson junction array shown in
figure~\ref{schematic}, and to obtain an explicit
expression for the key parameters $\kappa$ and $\delta$.  Our
approach follows closely that of reference \cite{Swi&al92}, which
treated the special case of a pure resistive load.  Starting from the
full circuit equations, we apply a first-order averaging method which
is valid in the weakly-coupled limit, but holds for a general
inductor-resistor-capacitor (LRC) load.  We also add some new
observations about the behavior of the averaged system.  A nice
feature of the present analysis is that the results admit a direct
physical interpretation:  The combined current of the Josephson
junction oscillators acts as a periodic driving voltage for the LRC
circuit, and {\em the in-phase oscillation is stable when the
Josephson junction frequency is larger than the resonant frequency
of the LRC circuit. } Conversely, if the junction frequency is smaller
than the resonant frequency of the load, then the manifold of
incoherent states is stable.  Thus we recover (and refine) an oft
quoted piece of conventional wisdom found in the
Josephson array literature, that the in-phase state is
stable for inductive loads, but unstable for capacitive loads.
Furthermore, we observe that the way to extract the most energy
from the Josephson junctions is to tune their frequency to match the
resonant frequency of the LRC circuit.

Strogatz and Mirollo~\cite{Str&Mir93} computed analytically the
stablility of the splay-phase states (the most symmetric of the
incoherent states) of the unaveraged system in the $N \rightarrow
\infty$ limit.  Surprisingly, their results showed excellent
agreement with numerical calculations~\cite{Nic&Wie92} even for
$N$ as small as four.  Though Strogatz and Mirollo did not assume
weak coupling, their stability results do not easily allow a physical
interpretation.  We show that our results derived from the averaged
system (for any N) agrees with their results (for $N \rightarrow
\infty$) in the weak coupling limit.  The weak coupling limit also
provides us with a simple physical interpretation of these results.

\section{Derivation of the averaged equations}
\label{sec:averaging}

Consider the array depicted in fig. \ref{schematic}.  The goal of this
section is to show that, in the limit of large shunt impedence, the
circuit dynamics is governed by differential equations of the form
(\ref{target}), and to evaluate $\kappa$ and $\delta$ in terms of the
physical parameters of the system.

Our starting point is the Kirchoff equations for the circuit.  We
assume that the $N$ junctions are identical and have negligible
capacitance ($\beta = 0$ in the common notation).
The governing circuit equations are

\begin{equation}
\frac{\hbar}{2eR_J} \dot{\phi}_k + I_c \sin \phi_k + \dot{Q} = I_b
\label{Kirkhoff1}
\end{equation}
\begin{equation}
L \ddot{Q} + R \dot{Q} + \frac{Q}{C} =
\frac{\hbar}{2e} \sum_{j=1}^{N} \dot{\phi}_j
\label{Kirkhoff2}
\end{equation}
where $k$=1, 2, . . . , $N$.  Here, $\phi_k$ is the quantum phase
difference across the $k^{th}$ Josephson junction, $R_J$ is the
junction resistance, $I_c$ is the junction critical current,  $Q$ is the
charge on the load capacitor, $I_b$ is the applied bias current,  $L$,
$R$, and $C$ are the load inductance, resistance, and capacitance,
respectively; $\hbar$ is Planck's constant divided by 2$\pi$, $e$ is
the electron charge, and the overdot denotes differentiation with
respect to time $t$.  Substitution
of Eq. (\ref{Kirkhoff1}) into Eq. (\ref{Kirkhoff2}) yields

\begin{equation}
L \ddot{Q} + (R+NR_J) \dot{Q} + \frac{Q}{C} =
NR_JI_b - R_JI_c \sum_{j=1}^{N} \sin\phi_j
\end{equation}
It is convenient to shift the load variable $Q$ by a constant

\begin{equation}
\frac{Q}{C} - NR_JI_b \rightarrow \frac{Q}{C}
\end{equation}
so that this becomes

\begin{equation}
L \ddot{Q} + (R+NR_J) \dot{Q} + \frac{Q}{C} =
 - R_JI_c \sum_{j=1}^{N} \sin\phi_j
\label{loadeqn}
\end{equation}

In order to compare arrays having different numbers of
junctions, it is natural to define scaled load parameters:

\begin{equation}
l=L/N; \: r=R/N; \: c=NC
\label{scaledload}
\end{equation}

Introducing the dimensionless time $\tau$ and charge $q$
defined by

\begin{equation}
\tau = \frac{2e}{\hbar}R_JI_ct
\label{tauDef}
\end{equation}
\begin{equation}
q = lR_JI_c \left(\frac{2e}{\hbar}\right)^{2}Q
\end{equation}
the circuit equations (\ref{Kirkhoff1}) and (\ref{loadeqn})
become

\begin{equation}
\dot{\phi}_k + \sin \phi_k + \epsilon \dot{q} = \alpha
\label{simpleI}
\end{equation}
\begin{equation}
\ddot{q} + \gamma \dot{q} + {\omega_0}^{2} q
= - \frac{1}{N} \sum_{j=1}^{N} \sin\phi_j
\label{simpleV}
\end{equation}
where the overdot now denotes differentiation
with respect to dimensionless time $\tau$, and where

\begin{equation}
\epsilon =      \frac{\hbar}{2eI_cl}
\label{epsilon}
\end{equation} \begin{equation}
\alpha=I_b/I_c
\end{equation} \begin{equation}
\gamma = \frac{(r+R_J)\hbar}{2eR_JlI_c}
\label{gamma}
\end{equation} \begin{equation}
{\omega_0}^{2} = \frac{1}{lc} \left(\frac{\hbar}{2eR_JI_c}\right)^{2}
\label{omega0}
\end{equation}
Note that $\gamma$ includes the effect of both the junction
resistance and the load resistance.  Thus $\gamma$ is never zero: in fact
$\gamma \ge \epsilon$ with equality when the load resistance is zero.

Up to this point, Eqs. (\ref{simpleI}), (\ref{simpleV}) are merely
scaled versions of the exact circuit equations (\ref{Kirkhoff1}),
(\ref{Kirkhoff2}).  To get things in a form suitable
for averaging, we transform from the variables $\phi_k$ to natural
angles $\psi_k$ \cite{Swi&al92}.  The latter are ``natural" in the
sense that, in the uncoupled limit, the angular velocity
$\dot{\phi_k}$ is non-uniform, while $\dot{\psi_k}$ is a constant.
This is accomplished by the transformation \cite{Swi&al92}

\begin{equation}
\psi(\phi) = 2 \arctan \left[\sqrt{\frac{\alpha - 1}{\alpha + 1}}
\tan\left(\frac{\phi}{2} + \frac{\pi}{4}\right)\right]
\end{equation} \begin{equation}
\phi(\psi)= 2 \arctan \left[ \sqrt{\frac{\alpha + 1}{\alpha - 1}}
\tan\left(\frac{\psi}{2} \right)\right] - \frac{\pi}{2}
\label{phi-psi}
\end{equation}

Equation (\ref{simpleI}) becomes

\begin{equation}
\frac{1}{\sqrt{\alpha^{2}-1}} \dot{\psi_k} = 1 -
\frac{\epsilon \dot{q}}{\alpha-\sin(\phi_k)}
\label{oldtau}
\end{equation}
Note that $\omega = \sqrt{\alpha^{2}-1}$ is the frequency of an
uncoupled junction.  It is convenient to rescale time to units where
this frequency is unity, by taking
$\tau \sqrt{\alpha^{2}-1} \rightarrow \tau$, so that
Eqs.(\ref{oldtau}),(\ref{simpleV}) become

\begin{equation}
\dot{\psi_k} = 1 -
\frac{\epsilon \omega \dot{q}}{\alpha-\sin(\phi_k)}
\label{psidot}
\end{equation}
and
\begin{equation}
\omega^{2}\ddot{q} + \gamma \omega \dot{q} + {\omega_0}^{2} q
= -\frac{1}{N} \sum_{j=1}^{N} \sin\phi(\psi_j)
\label{qdotdot}
\end{equation}

The system of equations (\ref{psidot}, \ref{qdotdot}) is exact, and so
far we have not made an assumption of weak coupling.  Notice that
Eq.(\ref{psidot}) is of the form $\dot{\psi_k}=1+O(\epsilon)$.  For
small $\epsilon$, we can obtain the drift of $\psi_k$ by averaging
Eq.(\ref{psidot}) over one period of the oscillation to obtain

\begin{equation}
\langle \dot{\psi_k} \rangle = 1 - \frac{1}{2\pi}\int_{0}^{2\pi}
\frac{\epsilon
\omega \dot{q}}{\alpha - \sin\phi(\psi_k)} d\tau .
\label{avgpsidot}
\end{equation}
We can proceed by using the function $q(\tau)$ obtained by solving
Eqs.(\ref{psidot}) and (\ref{qdotdot}) with $\epsilon = 0$.
Mathematically, the resulting (first-order) averaged system
(\ref{avgpsidot}) is valid in the case where the Josephson junctions
are weakly coupled, i.e. when $\eps \ll 1$.  Physically,
Eq.(\ref{epsilon}) shows this occurs for sufficiently large values of
the load inductance (per junction), and a small current flows through
the load.

To find the appropriate expression for $q(\tau)$, we begin with the
$\epsilon = 0$ solution to Eq.(\ref{psidot}):

\begin{equation}
\psi_k(\tau) = \tau + c_k
\label{linearPsi}
\end{equation}
where the $c_k$ are arbitrary inital conditions.  From
Eq. (\ref{phi-psi}) there follows the useful trigonometric identity

\begin{equation}
\sin \phi(\psi) = \alpha -
\frac{\alpha^{2}-1}{\alpha-\cos\psi}
\label{unexpected}
\end{equation}
It is convenient to write this even function of $\psi$
in terms of its Fourier series

\begin{equation}
\sin\phi(\psi) =
\sum_{n=0}^{\infty}A_n \cos(n\psi)
\end{equation}
where, in particular,

\begin{equation}
A_1 = 2 ( \alpha^2 -1 - \alpha \sqrt{\alpha^2 -1} )
\label{coeffs.An}
\end{equation}
Note that $A_1$
is a decreasing function of $\alpha$
with $0 \ge A_1 > -1$ for $\alpha \ge 1$.

Combining Eqs.(\ref{qdotdot}), (\ref{linearPsi})-(\ref{coeffs.An})
yields

\begin{equation}
\omega^{2}\ddot{q} + \gamma \omega \dot{q} + {\omega_0}^{2} q
= -\frac{1}{N}\sum_{j=1}^{N}\sum_{n=0}^{\infty}
A_n\cos n(\tau+c_j)
\end{equation}
This equation has the steady state solution

\begin{equation}
q(\tau)=-\frac{1}{N}\sum_{j=1}^{N}\sum_{n=0}^{\infty}
B_n\cos(n(\tau+c_j)+\beta_n)
\label{q.tau}
\end{equation}
where
\begin{equation}
B_n^{2}=\frac{A_n^{2}}{
(n^{2}\omega^{2}-\omega_{0}^{2})^2
+ (\gamma n \omega )^{2}}
\label{Bn}
\end{equation}
\begin{equation}
\beta_n=\arctan\left[\frac{\gamma n\omega}
{ n^{2}\omega^{2}-\omega_{0}^{2} } \right]
\end{equation}
Note that $B_n$ and $\beta_n$ are just the well-known amplitude
and phase shift response of a linear damped oscillator driven at
frequency $n \om$.  The relative sign betweenthe correct branch of the inverse
tangent.  We will choose $B_n$ to be
positive, and since $A_1$ is negative, we have $0 \le \beta_1 \le \pi$.

The next step is substitute this expression for $q(\tau)$ back into
Eq.(\ref{avgpsidot}).  Note that the identity Eq.(\ref{unexpected})
allows us to rewrite Eq.(\ref{avgpsidot}) as

\begin{equation}
\langle \dot{\psi_k} \rangle = 1 - \frac{\epsilon
\omega}{2\pi}\int_{0}^{2\pi}
\dot{q}(\tau) \frac{\alpha-\cos(\tau+c_k)}{\alpha^{2} - 1}
d\tau
\label{avgpsidot2}
\end{equation}
Now, since $q(\tau)$ is $2\pi$-periodic, it is evident from
Eq.(\ref{avgpsidot2}) that only the fundamental Fourier component
of $\dot{q}(\tau)$ contributes to the integral, with the result

\begin{equation}
\langle \dot{\psi_k} \rangle = 1 + \frac{\epsilon B_1}{2N\omega
}\sum_{j=1}^{N}
\sin \left(c_j - c_k - \beta_1 \right)
\label{avgpsidot3}
\end{equation}
The final step is to replace the ``initial values" $c_k$ by their
slowly evolving counterparts $\langle \psi_k(\tau) \rangle$, and
drop the angular brackets to get the first-order averaged equations

\begin{equation}
\dot{\psi_k} = 1 + \frac{\epsilon B_1}{2N\omega }\sum_{j=1}^{N}
\cos \left(\psi_j - \psi_k - \delta \right)
\label{mainresult}
\end{equation}
where $\delta=\frac{\pi}{2}+\beta_1$ is given by
\begin{equation}
\sin \delta = \frac{\om^2 - \om_0^2}{\sqrt{(\om^2 - \om_0^2)^2 + (\gamma
\om)^2}}
\label{delta}
\end{equation}
with $\frac{\pi}{2} \le \delta \le \frac{3\pi}{2}$.  This is the main
result of the paper.  Note that $\del$ is simply related to $\beta_1$,
which is the phase shift of the linear LRC circuit
(Eq.(\ref{simpleI})) driven at the frequency ($\om$) of the
uncoupled Josephson Junctions (Eq.(\ref{simpleV})).

In terms of the parameters in the original equations,
(\ref{Kirkhoff1}) and (\ref{Kirkhoff2}),
the important dimensionless frequencies are:
\begin{equation}
{\omega_0}^{2} = \frac{1}{LC} \left(\frac{\hbar}{2eR_JI_c}\right)^{2}
{\omega}^2 = \left(\frac{I_b}{I_c}\right)^2 -1.
\end{equation}

\section{In-phase and incoherent states}

We start this section by recalling some results of references
\cite{Ash&Swi92}, \cite{Swi&al92}, and \cite{Wat&Str94}.
Equation (\ref{mainresult}) admits two types of solutons:
The {\em in-phase} (or coherent)
solution, where all of the angles are equal, and {\em incoherent}
solutions, where the ``center of mass" of the $N$ angles $\psi_i$,
when they are placed on a circle, is at the center of the circle.  The
incoherent solutons rotate rigidly and have period exactly $2 \pi$. (It
is easy to see that the coupling terms in equation (\ref{mainresult})
cancel out for incoherent solutions.)

The in-phase solution is unique: In geometric terms, this periodic
orbit is a circle (a one-dimensional manifold)
in the $N+2$ dimensional phase space.
It is natural to ask ``How many incoherent solutions are there?''
There is a unique incoherent solution if $N=2$ or $3$, but
an infinite number of incoherent solutions if $N > 3$.
In fact the set of incoherent solutions is an
$N-2$ dimensional manifold for any $N \ge 3$, called the
{\em incoherent manifold} by Watanabe and Strogatz
\cite{Wat&Str94}.
The incoherent manifold is foliated by circles, which are the the
incoherent
solutions.

We can compute the dimension of the incoherent maifold
as follows:
Place $N-2$ oscillators on the circle of radius 1, so that their center of
mass is not at the origin.  (This gives the $N-2$ dimensions of the
incoherent manifold.)
Then, the position of the last 2 oscillators is uniquely determined
since the
center of mass of all $N$ oscillators is at the origin.
Note that if the center of mass of the first $N-2$ oscillators is
farther than $2/(N-2)$ from the origin, then it is not possible to place
that
last two oscillators so that
the center of mass of all $N$ is at the origin.

A major result of reference \cite{Wat&Str94} is that
{\em unaveraged} systems with a ``sinusoidal''
nonlinearity, including Eqs. (\ref{Kirkhoff1}) and
(\ref{Kirkhoff2}),
have an incoherent manifold.  In other words, any initial condition in
the incoherent manifold is part of a periodic orbit.
Incoherent solutions can be defined for unaveraged systems in terms
of time delays\cite{Ash&Swi92,Ash&Swi93}.

The most symmetric of the incoherent solutions, with the angles all
equally spaced in time, is called the {\em splay-phase} solution.
The stability of the in-phase and splay-phase solutions in equation
(\ref{mainresult}) is easy to calculate, following reference
\cite{Ash&Swi92}(section 6). We give the stability of these periodic
solutions in terms of the Floquet exponents, which are analogous to
the eigenvalues of the linearization about a fixed point. A given
periodic orbit has as many exponents as there are phase space
dimensions; if any of the exponents has positive real part, then a
typical perturbation will grow exponentially, and the periodic orbit is
unstable.  All periodic orbits have at least one zero Floquet exponent
corresponding to a perturbation along the orbit.

The in-phase solution has a single Floquet exponent equal to zero,
and $N-1$ exponents equal to
\begin{equation}
\lambda_{in-phase} = \frac{- \eps B_1}{2 \omega} \sin\delta
\end{equation}
where $\delta$ is given by Eq.(\ref{delta}) and $B_1$ is
\begin{equation}
B_1=\frac{2\omega(\sqrt{\omega^2+1}-
\omega)}{\sqrt{(\omega^{2}-\omega_{0}^{2})^2
+ (\gamma \omega)^{2}}}
\end{equation}
from Eqs.(\ref{coeffs.An}) and (\ref{Bn}).
The splay-phase solution has $N-2$ Floquet exponents equal to zero
and a complex conjugate pair
\begin{equation}
\label{lamsplay}
\lambda_{splay} = \frac{\eps B_1}{4 \omega}(\sin \delta \pm
i \cos\delta)
\end{equation}
We see the crucial role of $\delta$. If $\sin\delta > 0$ then the in-
phase
solution is stable and the splay-phase is unstable.  If $\sin\delta < 0$
then in-phase solution is unstable and the spay-phase solution is
neutrally stable.  Watanabe and Strogatz \cite{Wat&Str94} showed
that every incoherent solution of the averaged equation
(\ref{mainresult}) has the same stability as the
splay-phase solution.  (The sign of each Floquet exponent is the same,
although the magnitude varies.)  An incoherent solution is neutrally
stable to the $N-3$ perturbations which leaves it in the incoherent
manifold.  If $\sin\delta < 0$ then perturbations normal to the
incoherent manifold decay away, and we say that the incoherent
manifold is stable.  The incoherent manifold is stable exactly when
the splay-phase state is neutrally stable.

{}From Eq.(\ref{delta}) we see that the in-phase solution is
stable when
$\omega > \omega_0$, and the incoherent manifold is
stable when
$\omega < \omega_0$.  There is no bistability in the averaged
equations.
We note that numerical simulations of the unaveraged array
equations show bistability
in an LC-shunted Josephson array \cite{Had&Bea87,Tsa&Sch92}
(i.e. Equations (\ref{Kirkhoff1}) and (\ref{Kirkhoff2}) with $R=0$).
Therefore, the averaged equations have somewhat different
dynamics than the original system, though both have an incoherent
manifold of solutions.

Strogatz and Mirollo~\cite{Str&Mir93} computed the Floquet
exponents of the splay-state of Eqs.(\ref{Kirkhoff1}) and
(\ref{Kirkhoff2}) in the limit $N \rightarrow
\infty$.  They found that all but four of the Floquet exponents are
zero, (the fact that Watantabe and Strogatz later proved is true for
finite $N$), while the other four are the eigenvalues of a $4 \times 4$
matrix.  In our notation, the $4^{th}$ order characteristic equation is
\begin{equation}
\label{characteristic}
\lam^4 + \gamma \lam^3 + ({\omega_0}^2 + \omega^2) \lam^2
+ [ \eps\omega ( \sqrt{\om^2 + 1} - \om ) + \gamma \om^2 ] \lam
+ {\om_0}^2 \om^2 = 0
\end{equation}
These eigenvalues derived for the case $N \rightarrow
\infty$ are in excellent agreement with numerical
calculations~\cite{Nic&Wie92} even for $N=4$, and it is conceivable
that the result is exactly true for any $N$.  In any event, it makes
sense to compare this formula with the eigenvalues derived from the
averaged system, as we now do.

If we set $\eps = 0$, the charatcteristic equation
Eq.(\ref{characteristic}) factors
\begin{equation}
(\lam^2 + \gamma \lam + {\om_0}^2 ) ( \lam^2 + \om^2 ) = 0.
\end{equation}
which has a natural physical interpretation.  Namely,
the first factor corresponds to the decay of the current in
the LRC branch of the
circuit, while the pure imaginary eigenvalues $\pm i \om$ from the
second factor correspond to the oscillation frequency of a single
junction in the absence of a load.  If one
computes from Eq.(\ref{characteristic}) the order $\eps$ correction
to the Floquet exponents $\pm i \om$, one finds precisely the result
for the averaged system, namely Eq.(\ref{lamsplay}).  Physically,
then, the averaged result captures to lowest order the effect of
interactions between the load and the junctions.

Finally, we can see how accurately the averaged equations capture
the transition where the splay states go unstable.  In terms of our
own parameters, equation (14) of reference \cite{Str&Mir93} gives
the transition curve
\begin{equation}
{\om_0}^2 = \om^2 + \frac{\eps}{\gamma} \om (\sqrt{\om^2 + 1} -
\om)
\label{transition}
\end{equation}
Recalling that the averaged system has the corresponding transition
at $\om_0 = \om$, we see there is exact agreement only for $\eps =
0$.  However, for $\eps > 0$ the transition
curve determined from Eq.(\ref{transition}) never gets very far from
the diagonal when plotted on the $\om$-$\om_0$ plane, as shown in
fig.~\ref{plottransition}.
Note from Eqs.(\ref{epsilon}) and (\ref{gamma}) that
$\eps/\gamma = R_J/(r+R_J)$ so that this ratio never
gets too big:  $0 < \eps/\gamma \le 1$.  Moreover, for large
$\omega$ (i.e. the limit of large bias current $I_b$) the transition
curve always approaches the line $\omega=\omega_0$.

\section{Discussion}

Our main result is the derivation of the averaged
system (\ref{mainresult}) from the original
Eqs.(\ref{Kirkhoff1},\ref{Kirkhoff2}).  Watanabe and Strogatz
have shown that while the original equations are unusually tractable
owing to the existence of a great many constants of motion, the
averaged equations are completely solvable \cite{Wat&Str94}.  In
this paper we derived an explicit
formula for the coupling-phase $\delta$, which is the key parameter
governing the stability of both the in-phase and splay states.

An advantage of the averaged equations is that they admit a fairly
direct physical interpretation of the main features of the array
dynamics.  Usually, for a series
LRC combination one identifies two resonance frequencies, which we
can call the natural
resonance frequency $\omega_0$ and the shifted resonance
frequency $\omega_1$.  In
terms of our dimensionless units, we have

\begin{equation}
\omega_1 = \omega_0 - \gamma^2 /2
\end{equation}
where $\omega_0$ and $\gamma$ were defined earlier via
Eqs.(\ref{omega0}) and (\ref{gamma}), respectively.
In the presence of a periodic driving voltage at a frequency
$\omega$, the current oscillations (and thus the power dissipated in
the load) is greatest when $\omega=\omega_0$, while the capacitor's
charge oscillations are greatest when $\omega=\omega_1$.  Of
course, for ``high-Q" circuits, the current- and charge-response
curves
are sharply peaked and $\omega_0$ is very nearly equal to
$\omega_1$.

Now, we noted in the last section that the averaged Josephson array
equations display a single dynamical transition at $\omega =
\omega_0$.  It follows that {\em the maximum power delivered to
the LRC
load is attained at the transition point}.  We also recover an old
result found in the Josephson array literature
\cite{Jain&al84,Had89}, namely that stable in-phase operation of a
series array
requires that the load ``looks inductive":  an LRC load is said to look
inductive if its impedance has positive imaginary part, which is
equivalent to the condition $\omega > \omega_0$.  We note that the
averaged equations
admit an analogous stability principle for the splay state: the
splay state is stable if the load impedance has negative
imaginary part ({\em i.e.} if the load looks capacitive).

\newpage

\begin{center}
{\bf Figures}
\end{center}

\begin{figure}[htp]
\caption[schematic]{
\label{schematic}
Circuit schematic for the shunted Josephson array.
} \end{figure}

\begin{figure}[htp]
\caption[plottransition]{
\label{plottransition}
Bifurcation of the splay state as given by Eq.(\ref{transition}) for $\eps /
\gamma $ = 0, 0.25, 0.5, and 0.75, and 1.  The curve with
$\eps=0$ is the result for the averaged system.  The uppermost curve,
with $\eps / \gamma = 1$, corresponds to an LC load ($R=0$).
} \end{figure}

\end{document}